# 5G Wireless and Wired Convergence in a Passive Optical Network using UF-OFDM and GFDM

Colm Browning[1,*], Arman Farhang[2], Arsalan Saljoghei[1], Nicola Marchetti[2],
Vidak Vujicic[1], Linda E. Doyle[2] and Liam P. Barry[1]

[1]School of Electronic Engineering, Dublin City University, Glasnevin, Dublin 9, Ireland.
[2]CONNECT Research Centre, Dunlop Oriel House, Trinity College Dublin, Dublin 2, Ireland.
[*]colm.browning@dcu.ie

*Abstract*— The provision of both wireless and wired services in the optical access domain will be an important function for future passive optical networks (PON). With the emergence of 5[th] generation (5G) mobile communications, a move toward a dense deployment of small cell antenna sites, in conjunction with a cloud radio access network (C-RAN) architecture, is foreseen. This type of network architecture greatly increases the requirement for high capacity mobile fronthaul and backhaul links. An efficient way of achieving such connectivity is to make use of wavelength division multiplexed (WDM) PON infrastructure where wireless and wired services may be converged for distribution. In this work, for the first time, the convergence of 5G wireless candidate waveforms with a single-carrier wired signal is demonstrated in a PON. Three bands of universally filtered orthogonal frequency division multiplexing (UF-OFDM) and generalized frequency division multiplexing (GFDM), are transmitted at an intermediate frequency in conjunction with a digital 10Gb/s pulse amplitude modulation (PAM-4) signal in the downlink direction. Orthogonal frequency division multiplexing (OFDM) is also evaluated as a benchmark. Results show, for each waveform, how performance varies due to the 5G channel spacing - indicating UF-OFDM's superiority in terms of PON convergence. Successful transmission over 25km of fibre is also demonstrated for all waveforms.

## I. Introduction

With the onset of the internet of things (IoT), and the continued increase in demand for high speed streaming services, it is imperative that mobile networks are augmented in order to provide higher speeds and increased flexibility, to a greater number of users. To this end, there has been much debate over which waveform can efficiently meet these needs, and eventually be implemented in 5[th] generation (5G) mobile communications [1]. The exploration of the suitability and performance of various contending waveforms candidates has so far been confined to the wireless domain.

Considering the ultra-dense (UD) deployment of small cell antenna sites that will be required to provide 1000× the aggregate data rate of 4G systems [1, 2], coupled with the C-RAN architectures, it follows that mobile backhauling (the delivery of wireless services from the network edge to wireless base-stations) and fronthauling (the delivery of services from centralised, consolidated baseband units (BBU) to remote radio heads (RRH)) will be a key requirement for 5G networks. Such an implementation places great importance on the fixed/optical portion of access networks, and the high capacity, low latency and flexibility [3] offered by PONs have made them an obvious choice to facilitate 5G development by providing optical backhaul and fronthaul of wireless signals [1, 4], in a cost effective manner [5]. It is clear that 5G candidate waveforms must be studied, not only in the wireless domain, but also in the optical domain where their suitability for transmission through optical access networks, and their potential for integration alongside other services – wireless-wired convergence – must be evaluated. In this work, for the first time, we demonstrate the downlink transmission of 5G candidates, UF-OFDM and GFDM, converged with a single-carrier wired signal in a PON.

Both UF-OFDM and GFDM are considered to be candidate waveforms for 5G wireless networks [6]. Like OFDM, they are multicarrier modulation schemes and are digitally implemented using an (inverse) fast Fourier transform ((I)FFT). Both are described in more detail in section II but, generally, the key differences between these waveforms and OFDM is that they utilise filtering techniques, at the subcarrier and/or resource block level, to modify the spectral properties of their signals, leading to lower out-of-band (OOB) emissions compared to OFDM. UF-OFDM is a modified version of OFDM which limits the OOB emissions of its subcarriers - and hence becomes more robust to the synchronization errors - by employing linear filtering. This brings filter transient periods to the each UF-OFDM symbol, known as the signal ramp-up and ramp-down durations [7]. To avoid the filter transients in GFDM signals, circular pulse-shaping is deployed where a block of time symbols are circularly convolved with the GFDM prototype filter. GFDM uses one cyclic prefix (CP) for a block of its time symbols to absorb the channel transient response and thus ease the channel equalization procedure. This makes GFDM more bandwidth efficient than UF-OFDM and OFDM [6].

Previous work has shown the convergence of single-carrier wired services with 4G LTE signals [5, 8], but 5G signals have not been studied before in this context. The use of UF-OFDM to provide multiple wired/wireless services has previously been demonstrated [9] but this system architecture does not represent a straightforward augmentation of current PON standards as it would involve the implementation of a new multicarrier access service. In this work, we propose, and experimentally demonstrate, a converged 5G PON system which utilises a 4-level PAM (PAM-4) signal as the wired service. PAM-4 was chosen as it has recently gained interest for use in future optical access networks as its low complexity, and compatibility with current intensity modulation/direct detection (IM/DD) implementations, make it a cost-effective solution for future PONs [10, 11]. The 5.5GBaud (10Gb/s plus FEC accommodation) baseband PAM-4 signal is transmitted in the downlink direction in tandem with three 5G wireless bands. Transmission with OFDM in place of the 5G bands is also demonstrated for system comparisons. The results presented in this work show the successful transmission of the converged wired and wireless signals, as well as the advantages offered by UF-OFDM and GFDM over OFDM, in terms of performance in the presence of the wired signal, and for reduced guard-bands between the wireless channels.

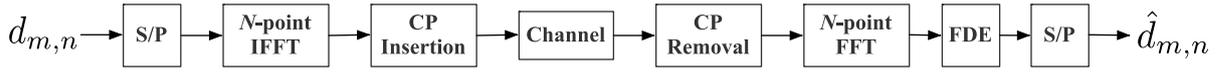

Fig. 1(a): Baseband structure of OFDM.

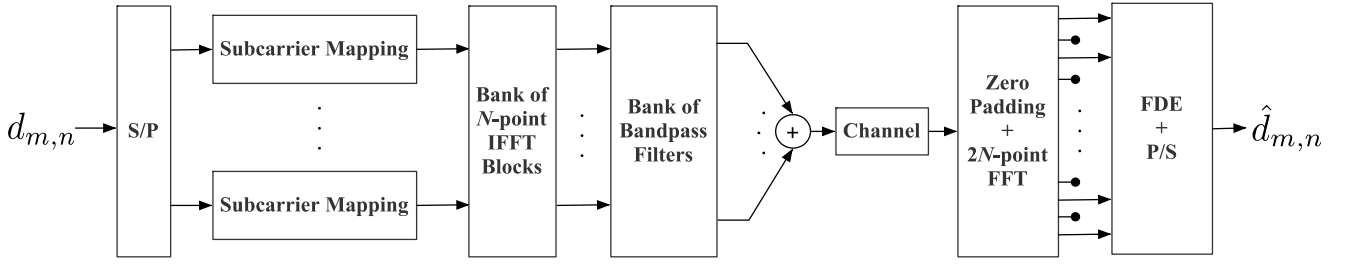

Fig. 1(b): Baseband structure of UF-OFDM.

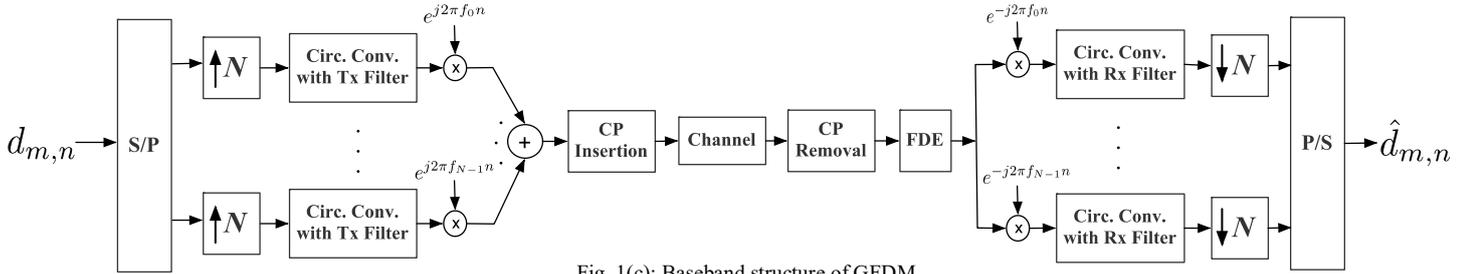

Fig. 1(c): Baseband structure of GFDM.

Fig. 1. Baseband system models for (a) OFDM, (b) UF-OFDM, and (c) GFDM.

## II. OFDM, UF-OFDM AND GFDM

### A. OFDM

To form the *m-th* OFDM symbol, the QAM data symbols in the vector $\boldsymbol{d_m} = [d_{m,0}, \ldots, d_{m,N-1}]^T$ are modulated through an *N*-point IFFT operation. This is equivalent to the summation of *N* coefficients, each scaled by the data symbols $d_{m,n}$. These tones are centred at the frequencies $f \in \{0, \frac{1}{T}, \ldots, \frac{N-1}{T}\}$ where *T* is the OFDM symbol duration. After forming a given OFDM symbol *m*, $N_{CP}$ samples from the end of the symbol are appended at the beginning as a cyclic prefix (CP) to absorb the channel transient response. The presence of a CP which is longer than the channel impulse response (CIR) converts the linear convolution of the transmit signal with the channel into a circular convolution, after CP removal at the receiver. Hence, the effect of the channel can be equalized through a single-tap equalization in the frequency domain, i.e. frequency domain equalization (FDE). Thus, the estimate of the transmitted symbols $\hat{d}_{m,n}$ can be obtained after the parallel to serial conversion. The baseband OFDM system model is presented in Fig. 1 (a).

Even though the CP simplifies channel equalization, it adds overhead to the signal as the transmission of *M* OFDM symbols imposes an overhead with the length equal to $MN_{CP}$. This leads to a bandwidth efficiency loss of $\frac{N_{CP}}{N}$. Moreover, OFDM suffers from a large amount of OOB emissions leading to additional bandwidth efficiency loss as large guard-bands are needed for the aggregation of different OFDM signals in the frequency domain.

### B. UF-OFDM

In UF-OFDM, the whole available bandwidth is split into different sub-bands akin to the physical resource blocks (PRBs) in LTE systems. To tackle the OOB emissions as well as the in-band leakage problems of OFDM, the aforementioned sub-bands are linearly filtered with the modulated version of a Dolph-Chebyshev bandpass filter, [6, 7]. Performing this filtering at the PRB level allows for a reduction in filter length, so that the overhead in UF-OFDM is no more than that required for the CP of an equivalent OFDM signal. Consequently, UF-OFDM overhead remains the same as in OFDM, while the spectral containment is enhanced [7]. The baseband system model of UF-OFDM is shown in Fig. 1 (b).

UF-OFDM modulation in the baseband can be summarized into three steps; (i) mapping the QAM data symbols to the allocated sub-bands to a given user, (ii) $N$-point IFFT operation per sub-band while the subcarriers at the position of the remaining sub-bands are set to zero, (iii) bandpass filtering operation per sub-band, and superposition of the signals relating to all the sub-bands to form the UF-OFDM transmit signal. After the signal is passed through the channel, the position of the receiver time window is identified through the synchronisation procedure. Then the UF-OFDM signal demodulation is performed in two steps, given that the channel length is $L$; (i) zero padding the $N+L-1$ received signal samples at a given symbol $m$ to the length $2N$, and performing a $2N$-point FFT operation, (ii) given perfect knowledge of the channel response, the channel distortions can be equalised in the same manner as OFDM through a single-tap equalisation operation on the odd output bins of the FFT block. As it is pointed out in [6], processing the odd FFT outputs while neglecting the even outputs is simply a frequency decimation operation. Hence, to simplify the UF-OFDM receiver structure and reduce its computational cost, time aliasing, as an equivalent of the frequency decimation, can be performed before feeding the signal to the FFT block. Consequently, we can perform an $N$-point FFT operation along with the FDE at the receiver, i.e. the same procedure typically employed at an OFDM receiver.

## C. GFDM

GFDM is a block-based modulation scheme where a time-frequency block of $M$ time-symbols and $N$ subcarriers form one GFDM block, [13]. Similar to OFDM and UF-OFDM, we consider the subcarrier spacing of GFDM equal to $1/T$. An appealing property of GFDM is that it can handle the channel transient response using only one CP for $M$ symbols and hence reduce the signal overhead. Moreover, GFDM localises each subcarrier in the frequency domain through its so-called circular pulse-shaping/filtering procedure while completely removing the filter transients – an operation known as 'tail-biting' [13]. Filtering each subcarrier in this manner has led to GFDM being referred to as non-orthogonal as it minimises overlapping between subcarriers, and hence increases GFDM's tolerance to timing synchronisation errors [12]. However, it has been shown that GFDM suffers from relatively high OOB emissions. This is due to the fact that similar to OFDM, GFDM is based on the transmission of a number of pure tones truncated with a rectangular window [6]. The difference between OFDM and GFDM is that while in OFDM, one QAM symbol modulates a single tone, a QAM symbol in GFDM modulates multiple tones. More details about GFDM derivation and its relation with OFDM can be found in [6]. Another drawback of GFDM might be the latency that is imposed by the transmission of a block of symbols making symbol by symbol detection impossible. Consequently, large values of symbols in a GFDM block may not be reasonable, as apart from imposing a long latency, long blocks might go through time variation of the channel and thus experience severe performance degradation.

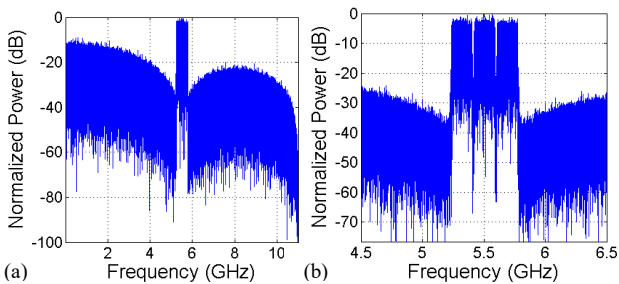

Fig. 2: (a) Spectrum of three UF-OFDM bands converged with a 5.5GBaud PAM-4 and (b) a zoomed version of the same spectrum.

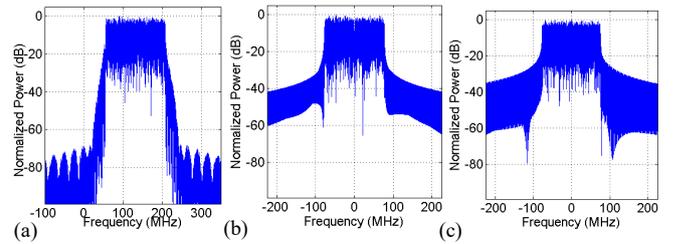

Fig. 3: Baseband spectra of single bands of (a) UF-OFDM, (b) GFDM and (c) OFDM.

The equivalent baseband system model of GFDM is depicted in Fig. 1 (c). GFDM modulation is performed in four stages; (i) upsampling the GFDM time-symbols by a factor of $N$, (ii) circular convolution of the upsampled symbols with the transmitter prototype filter to perform the circular pulse-shaping, (iii) upconversion of a given subcarrier $n$ to its corresponding subcarrier centre frequency $f_n = \frac{n}{T}$, (iv) superposition of all the subcarrier signals and insertion of a CP longer than the CIR. After a given GFDM block is received at the receiver, the CP is removed and the channel distortions can be compensated through an FDE operation similar to OFDM and UF-OFDM. After the channel is equalised, GFDM demodulation can be performed in three stages to estimate the transmitted symbols $\hat{d}_{m,n}$; (i) downconversion of each subcarrier to baseband, (ii) circular convolution of the resulting signal with the prototype filter at the receiver[1], and (iii) downsampling the circular convolution output by a factor of $N$. It is worth noting that, in this work, for a low complexity implementation of GFDM, we deploy the modem structure that is proposed in [14] while using the zero-forcing (ZF) receiver. It is known that GFDM suffers from some performance penalty compared with OFDM which is due to the non-orthogonality of its subcarriers [13, 14].

TABLE I: MULTICARRIER PROPERTIES

| | Multicarrier Properties | | |
|---|---|---|---|
| | *UF-OFDM* | *GFDM* | *OFDM* |
| IFFT size | 1024 | 1024 | 1024 |
| Symbol Rate (MHz) | 1.95 | 1.95 | 1.95 |
| Subcarriers | 78 | 78 | 78 |
| Sub-bands | 13 | n/a | n/a |
| Overlapping factor | n/a | 5 | n/a |
| Filter Type | Dolph-Chebychev FIR | PHYDYAS | n/a |
| Filter Length (Sa) | 74 | 9 | n/a |
| Cyclic Prefix (%) | 0 | 0.625 | 3.125 |
| Raw Data Rate per 5G band (Gb/s) | 0.61 | 0.61 | 0.61 |

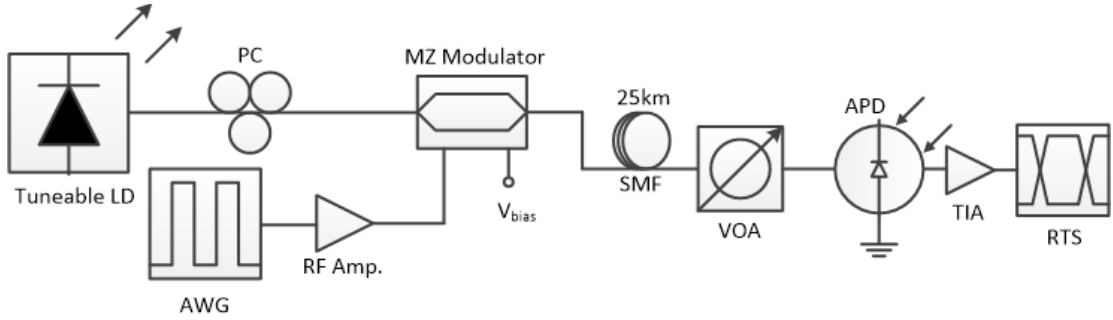

Fig. 4: Experimental PON setup.

### III. EXPERIMENTAL SETUP

#### A. Signal Generation

UF-OFDM, GFDM, OFDM and PAM signals were generated using Matlab. Although the overheads required varied for each multicarrier waveform, a similar bandwidth was maintained for fair comparison. In all cases, 78 subcarriers were modulated using 16 quadrature amplitude modulation (16-QAM) at a subcarrier symbol rate of 1.95MHz giving a raw data rate of 0.61Gb/s, and bandwidth of 152MHz per 5G band. In all cases the multicarrier waveforms were hard clipped to 80% of their original maxima resulting in reduced peak-to-average power ratios (PAPR) which varied between 11-12dB. Specific properties of each waveform can be found in table 1.

For each multicarrier waveform, three 0.61Gb/s 5G bands were generated using differing data streams. The bands were upsampled, modulated onto different intermediate frequencies and digitally added to the PAM waveform. The central band was placed in the spectral gap between the main lobe and the side-lobe of the 5.5GBaud PAM-4 signal (5.5GHz). The higher and lower frequency bands were placed adjacent to the central band, at intermediate frequencies which varied according to the desired guard-band between the three 5G bands. Fig. 2(a) shows an example composite spectrum while Fig. 2(b) is a zoomed version of the same transmitted composite spectrum, showing three UF-OFDM bands centred at 5.5GHz with a 15MHz guard-band between each band. At the digital receiver, the multicarrier signals are resampled and a 12th order Gaussian filter is used to extract each band for processing. For the PAM-4 signal, the adaptive equalizer was a 13-tap finite impulse response (FIR) filter and the tap weights were updated according to a decision-directed least-mean square (DD-LMS) algorithm [15].

Fig. 3 shows the baseband versions of a single multicarrier band. The OOB emission characteristic of each waveform can be clearly observed. In Fig. 3(a) the effect of the sub-band linear filtering used with UF-OFDM is apparent as OOB emission is highly reduced compared to GFDM and OFDM.

#### B. PON Setup

The experimental setup is presented in Fig. 4. The composite 5G/PAM-4 signal was loaded into the arbitrary waveform generator (AWG) which operated at 20GSa/s. The signal was amplified and used to drive a Mach-Zehnder modulator (MZM), biased at quadrature, which modulated the light from a tuneable laser diode (TLD), via a polarization controller (PC). The input power to the 25km of single mode fibre (SMF) was 2dBm. A variable optical attenuator (VOA) was used to control the input optical power to the avalanche photodiode (APD), with integrated trans-impedance amplifier (TIA), which exhibited saturation close to -14dBm. The received signal was sampled at 50GSa/s by a real time oscilloscope (RTS). Resampling, synchronisation, filtering, channel estimation/equalization, error vector magnitude (EVM) and bit error rate (BER) analysis were performed offline.

### IV. RESULTS AND DISCUSSION

In this type of hybrid PON system where multiple services may be delivered over a single fibre, spectral containment of the wireless signals is of high importance, not only in order to maximise spectral efficiency, but also to lessen the potential mutual impact of/on neighbouring wired services. Table II shows the EVM of each received 5G band under a variety of conditions. In all cases, the signals were transmitted over 25km of SMF, the received optical power was -14dBm and the measured BER of the PAM signal was below $1\times10^{-4}$. The intermediate frequency of the central band (band 2) is set at 5.5GHz and an initial spectral guard-band of 15MHz (10% of the 5G bandwidth as is common in 4G transmission [16]) is set between the wireless bands. The table shows the performance when the guard-band is reduced to 10MHz and 5MHz by varying the intermediate frequencies of the outer bands (bands 1 and 3). For comparison, EVMs are presented where the three 5G bands are transmitted without PAM (15GHz guard-band) and also when only the central band is transmitted. Results in table II show how UF-OFDM and OFDM exhibit similar performance when only a single wireless band at 5.5GHz is transmitted as UF-OFDM's lower OOB emission property does not have any bearing on system performance in this case. For the same condition, GFDM exhibits a penalty (~1% EVM) compared to UF-OFDM and OFDM and this is attributed to noise enhancement caused by the non-orthogonality of the GFDM subcarriers, as outlined in section II.C. As the wired PAM-4 signal is added, and the wireless guard-band decreased, the performance improvement due to reduced OOB emission is evident for GFDM, and particularly UF-OFDM, compared to OFDM. The results clearly indicate the superiority of UF-OFDM, as is to be expected given the spectral profiles in Fig. 3. Interestingly, for reduced guard-bands, GFDM exhibits performance in between that of OFDM and UF-OFDM, while offering higher tolerance to timing synchronisation errors, as well as reduced overhead, compared to *both* OFDM and UF-OFDM. Differences in the performances of each band are caused by the differing levels of interference they experience from the neighbouring wireless and/or PAM signals, as well as the slightly different PAPR of each band. The photo-receiver also exhibits a frequency roll-off of ~3dB after 5.5GHz.

---
[1] The prototype filters at the GFDM transmitter and receiver are different. This is due to the fact that GFDM is a non-orthogonal waveform and matched filter in this case is not the optimal choice.

TABLE II: 5G PERFORMANCE FOR VARIOUS CHANNEL SPACINGS

| EVM (%) | UF-OFDM | | |
|---|---|---|---|
| | *Band 1* | *Band 2* | *Band 3* |
| Single band w/o PAM | | 5.52 | |
| 15MHz w/o PAM | 5.93 | 6.32 | 6.35 |
| 15MHz w/ PAM | 6.05 | 6.38 | 6.57 |
| 10MHz w/ PAM | 6.27 | 6.39 | 6.7 |
| 5MHz w/ PAM | **6.51** | 6.68 | 6.77 |
| | GFDM | | |
| | *Band 1* | *Band 2* | *Band 3* |
| Single band w/o PAM | | 6.5 | |
| 15MHz w/o PAM | 7.01 | 7.00 | 7.43 |
| 15MHz w/ PAM | 7.59 | 7.15 | 8.11 |
| 10MHz w/ PAM | 7.62 | 7.77 | 8.73 |
| 5MHz w/ PAM | **8.37** | 8.47 | 9.95 |
| | OFDM | | |
| | *Band 1* | *Band 2* | *Band 3* |
| Single band w/o PAM | | 5.69 | |
| 15MHz w/o PAM | 6.96 | 7.03 | 7.16 |
| 15MHz w/ PAM | 7.46 | 7.38 | 7.78 |
| 10MHz w/ PAM | 9.28 | 11.71 | 10.27 |
| 5MHz w/ PAM | **10.33** | 12.79 | 10.25 |

Fig. 5 (a), (b) and (c) shows EVM per subcarrier, for the UF-OFDM, GFDM and OFDM central bands (band 2), respectively. The figure gives a clearer picture of the overall trends in each waveform as the guard-band is reduced. In all cases the outermost subcarriers experience degradation due to increased inter-band interference when the guard-band is reduced below 15MHz, however it is clear that less subcarriers are affected, and to a lesser degree, in the case of UF-OFDM compared to both GFDM, and OFDM – whose higher and lower frequency subcarriers are severely degraded for guard-bands of 10MHz and 5MHz. Again, the difference in degradation at either side of the band, is due to slight difference in PAPR of the neighbouring wireless bands – resulting in a small variance in interference levels.

Fig. 6 shows the EVM per subcarrier, for the lower frequency band (band 1), for all waveforms, when a 5MHz guard-band is employed. The inset shows the total received GFDM constellation, and the plotted lines are colour-coded with their corresponding average EVM values presented in table II. The figure shows how the higher frequency OFDM subcarriers (58 to 78) are impacted by the central wireless band. This trend decreases for GFDM. For the UF-OFDM signal, with the exception of the final 3 subcarriers, there is a slight improvement in performance as subcarrier frequencies increase. This is because lower frequency subcarriers are mildly impacted by the underlying PAM signal, whereas most of the higher frequency subcarriers experience little or no interference from the adjoining PAM or UF-OFDM signals. These trends are almost mirrored in Fig. 7 (also colour-coded with table II) which shows the higher frequency UF-OFDM, GFDM and OFDM band (band 3) where a guard-band of 10MHz is used. The inset shows the total received UF-OFDM constellation. Here, in the case of UF-OFDM, the influence of the PAM side-lobe can be observed at higher frequency subcarriers whereas lower frequency subcarriers experience very low levels of interference.

Fig. 8 shows BER versus received optical power relating to band 1 for each wireless waveform (15MHz guard-band), and also for the PAM-4 signal. The figure shows performance in the back-to-back (no fibre transmission) cases as well as transmission over 25km of fibre. The results are shown for received optical powers below -16dBm; at which points BER could be reliably calculated. In all cases there is no penalty due to fibre transmission, indicating the ability of GFDM's reduced CP, and UF-OFDM's multi-tap sub-band filtering, to effectively deal with dispersion in the optical domain. At these lower received powers (where Gaussian noise from the photo-receiver begins to dominate system performance) and with a 15MHz guard-band, UF- OFDM and OFDM display similar performances. At a BER of $1\times10^{-3}$ GFDM incurs a 1.2dB penalty in receiver sensitivity compared to OFDM and UF-OFDM due to the noise enhancement discussed previously. The inset in the figure shows the received eye diagram of the PAM-4 signal at -16dBm (BER = $2\times10^{-4}$). The 7% forward error correction (FEC) limit for PAM is set at $3.8\times10^{-3}$ indicating a required minimum received optical power of -18dBm for the system presented. Considering the optical launch power is 2dBm, this yields an optical power budget of 20dB based on the wired transmission alone, however this figure could be significantly improved by increasing the signal launch power. Of course, the system should operate so that satisfactory performance is attained for the wireless signals, taking into account additional operations or transmission which could be required after photo-detection. This would, in effect translate to a lower optical power budget. A way to achieve a 'balanced' performance would be to adjust the relative powers of the wired signal and the wireless bands [9], essentially trading off wired/wireless performance based on the precise needs of the system. For the experimental results shown in this work, the converged signals were set to have close to equal powers with the wired-to-wireless power ratio (WWPR) calculated as -1.36dB by integrating under the respective portions of the composite spectrum.

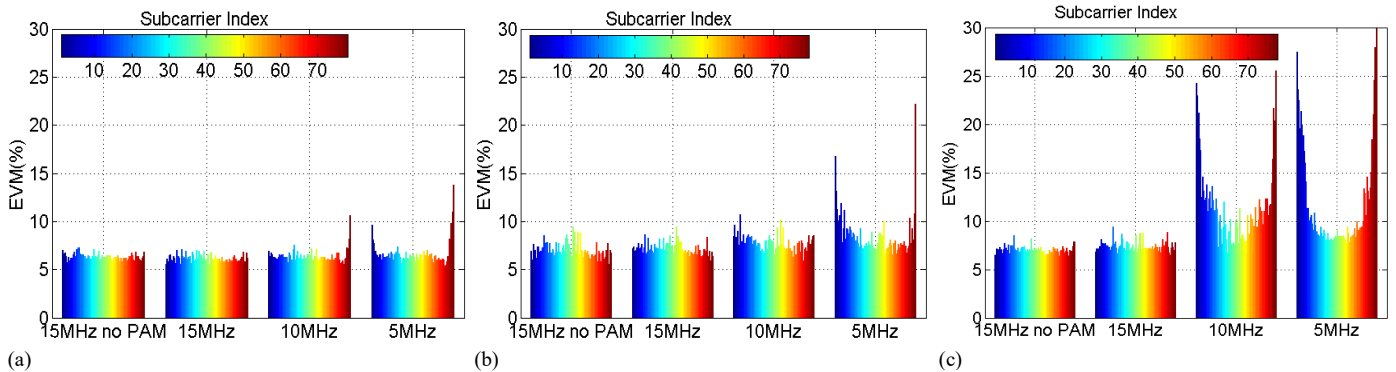

Fig. 5: EVM versus subcarrier index without the presence of PAM and for all guard-bands for (a) UF-OFDM, (b) GFDM and (c) OFDM.

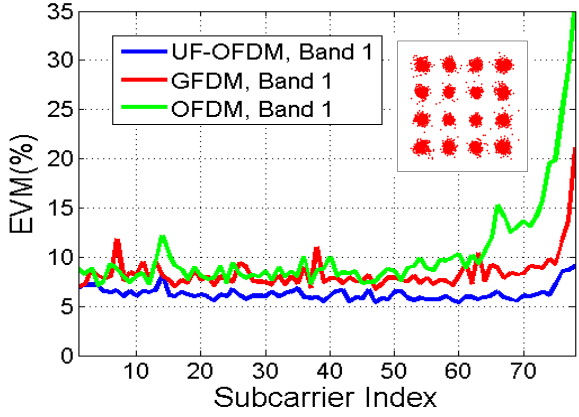

Fig. 6: EVM versus subcarrier index for band 1 of all waveforms with a 5MHz guard-band. Inset shows the total received GFDM constellation.

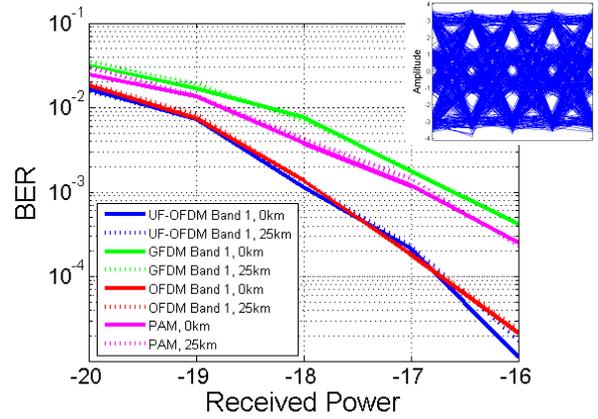

Fig. 8: BER versus received optical power for all wireless waveforms (15MHz guard-band) and for PAM-4, back-to-back and over 25km SMF.

Fig. 9 shows the performance of all three UF-OFDM bands with a 15MHz guard-band, in the back-to-back case as well as over 25km of SMF. There is a slight penalty incurred by the higher frequency band due to the frequency roll-off of the photo-receiver. Nevertheless, successful transmission is achieved without penalty due to fibre propagation.

IV. CONCLUSION

For the first time, the convergence of 5G candidate waveforms, UF-OFDM and GFDM, with a single-carrier wired service has been demonstrated in a PON. Transmission over 25km of SMF is performed without penalty, and EVMs of around 6% and 7%, for the converged UF-OFDM and GFDM services respectively, are achieved with a 10% channel guard-band (15MHz). Results also show how the use of these new waveforms can reduce the interference/channel spacing limitations posed by OFDM's high OOB emission, particularly in the case of UF-OFDM whose sub-band filtering properties allow for good performance in the case where a 5MHz guard band (~3% of total wireless channel bandwidth) is used between the wireless channels. For the same conditions it is shown that GFDM offers a reduced improvement (over OFDM) but in designing a next generation converged access networks, its ease of implementation, tolerance to synchronisation errors as well as increased bandwidth efficiency, compared to UF-OFDM, must be factored in.

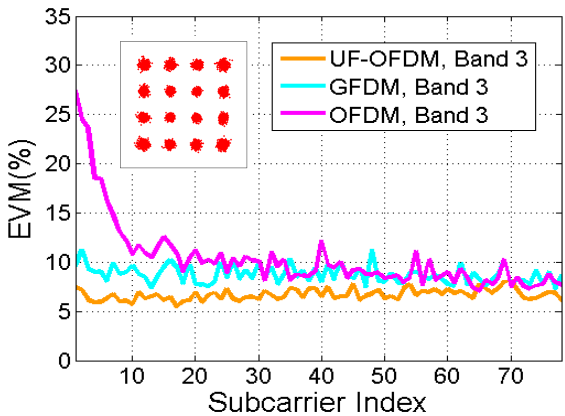

Fig. 7: EVM versus subcarrier index for band 3 of all waveforms with a 10MHz guard-band. Inset shows the total received UF-OFDM constellation.

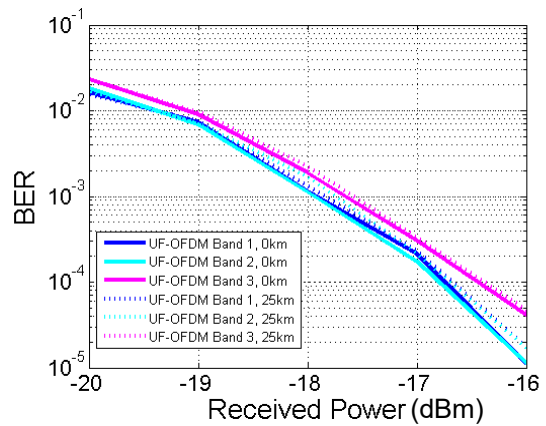

Fig. 9: BER versus received optical power for all UF-OFDM bands, with a 15MHz guard-band, in the back-to-back case as well as over 25km of SMF.

The convergence of these 5G wireless waveform candidates with a 10Gb/s PAM-4 wired signal, coupled with the expansion of available wavelengths in future WDM access, represents a straighforward augmentation of current PON technologies to enable the efficient development of 5G communications.

ACKNOWLEDGMENT

This work has been jointly supported through the SFI US-Ireland project (15/US-C2C/I3132), CONNECT (13/RC/2077), and IPIC (12/RC/2276) research grants.

REFERENCES

[1] J. Andrews et al., "What will 5G be?" IEEE J. Sel Areas Commu., vol 32, no 6, pp 1065-1082, June 2014.
[2] R.G. Stephen and R. Zhang, "Joint millimeter-wave fronthaul and OFDMA resource allocation in ultra-dense C-RAN," June 2016, arXiv:1603.09601v2 [cs.IT].
[3] S. Xu, S. Xu and Y. Tanaka, "Dynamic resource reallocation for 5G with OFDMA in multiple user MIMO RoF-WDM-PON," in proceedings Asia-Pacific Conference on Communications (APCC), pp. 480-484, Kyoto, 2015.
[4] X. Yu, J. B. Jensen, D. Zibar, C. Peucheret and I. T. Monroy, "Converged Wireless and Wireline Access System Based on Optical Phase Modulation for Both Radio-Over-Fiber and Baseband Signals," in IEEE Photon. Technol. Lttrs., vol. 20, no. 21, pp. 1814-1816, Nov. 2008.
[5] X. Hu, C. Ye and K. Zhang, "Converged mobile fronthaul and passive optical network based on hybrid analog-digital transmission scheme," in proceedings Optical Fiber Communications Conference (OFC), Anaheim, CA, 2016, pp. 1-3.
[6] B. Farhang-Boroujeny and H. Moradi, "OFDM inspired waveforms for 5G," IEEE Commun. Surveys Tuts. 2016.
[7] V. Vakilian, T. Wild, F. Schaich, S. ten Brink and J. F. Frigon, "Universal-filtered multi-carrier technique for wireless systems beyond LTE," 2013 IEEE Globecom Workshops (GC Wkshps), Atlanta, GA, 2013, pp. 223-228.
[8] A. Saljoghei, C. Browning and L. P. Barry, "Spectral shaping for hybrid wired/wireless PON with DC balanced encoding," in proceedings Microwave Photonics (MWP), pp. 307-310, Sendai, 2014.
[9] S. M. Kang, C. H. Kim, S. M. Jung and S. K. Han, "Timing-offset-tolerant universal-filtered multicarrier passive optical network for asynchronous multiservices-over-fiber," in IEEE/OSA Jnl. of Opt. Commun. and Netwrk., vol. 8, no. 4, pp. 229-237, April 2016.
[10] J. Gao, "Demonstration of the first 29dB Power Budget of 25Gb/s 4-PAM System without Optical Amlifier for Next Generation Access Network" in proceedings Optical Fiber Communications Conference (OFC), paper Th1I.2, Anaheim 2016.
[11] N. Argyris, I. Lazarou, S. Dris, P. Bakopoulos, C. Spatharakis, D. Kalavrouziotis, D. Apostolopoulos, and H. Avramopoulos, "Low cost 4-PAM heterodyne digital receiver for long reach passive optical networks," in proceedings OSA Advanced Photonics 2015, paper NeM4F.5, 2015.
[12] A. Aminjavaheri, A. Farhang, A. RezazadehReyhani and B. Farhang-Boroujeny, "Impact of timing and frequency offsets on multicarrier waveform candidates for 5G," Signal Processing and Signal Processing Education Workshop (SP/SPE), 2015 IEEE, Salt Lake City, UT, 2015, pp. 178-183.
[13] N. Michailow et al., "Generalized Frequency Division Multiplexing for 5th Generation Cellular Networks," in IEEE Transactions on Communications, vol. 62, no. 9, pp. 3045-3061, Sept. 2014.
[14] A. Farhang, N. Marchetti and L. E. Doyle, "Low-Complexity Modem Design for GFDM," in IEEE Transactions on Signal Processing, vol. 64, no. 6, pp. 1507-1518, March15, 2016.
[15] David Smalley, "Equalization Concepts: A Tutorial," Application Report, Texas Instruments, (1994).
[16] Nokia Siemens Networks, "Introducing LTE with maximum reuse of GSM assets", white paper, 2011, Available online [http://www.gsma.com/spectrum/wp-content/uploads/2012/03/lte1800mhzwhitepaper0.9.pdf].